\documentclass[10pt, conference]{IEEEtran}
\usepackage{ifpdf}
\usepackage{multirow}

\usepackage{color}

\usepackage{cite}
\ifCLASSINFOpdf
   \usepackage[pdftex]{graphicx}
\else
   \usepackage[dvips]{graphicx}
   \graphicspath{{../eps/}}
\fi
\usepackage[cmex10]{amsmath}
\usepackage{array}
\usepackage{mdwmath}
\usepackage{mdwtab}
\usepackage[font=footnotesize]{subfig}
\begin{document}
\title{Key Distribution Scheme without Deployment Knowledge}


\author{\IEEEauthorblockN{Pranav Agrawal, Joy Kuri}
\IEEEauthorblockA{Centre of Electronics Design and Technology\\
Indian Institute of Science\\
Bangalore, India\\
pagarwal,kuri@cedt.iisc.ernet.in}
}


%Telephone: (800) 555--1212, Fax: (888) 555--1212}


\IEEEoverridecommandlockouts
\IEEEpubid{9781--4244--3941--6/09/\$25.00~\copyright~2009 IEEE}

\maketitle

% \fi

\begin{abstract}
Many basic key distribution schemes specifically tuned 
to wireless sensor networks have been proposed in the literature. 
Recently, several researchers have proposed schemes in which they have used group-based 
deployment models and assumed predeployment knowledge of the expected locations of nodes. 
They have shown that these schemes achieve better performance than the basic schemes, 
in terms of connectivity, resilience against node capture and storage requirements. 
But in many situations expected locations of nodes are not available.
In this paper we propose a solution which uses the basic scheme, but does \emph{not} use 
group-based deployment model and  predeployment knowledge of the locations of nodes, and 
yet performs better than schemes which make the aforementioned assumptions. 

In our scheme, groups are formed \emph{after} deployment of sensor nodes, on the basis of 
their physical locations, and the nodes sample keys from disjoint key pools. 
Compromise of a node affects secure links with other nodes that are part of its group only. 
Because of this reason, our scheme performs better than the basic schemes and the schemes using 
predeployment knowledge, in terms of connectivity, storage requirement, and security. 
Moreover, the post-deployment key generation process completes sooner than in
schemes like LEAP+~\cite{leapjajodia}.

% Moreover, we show that in our scheme, the time required for key establishment
% is less than that in some other schemes.

\end{abstract}


\begin{IEEEkeywords}
Security; Key Distribution; Sensor Networks

\end{IEEEkeywords}

% \fi

\IEEEpeerreviewmaketitle
\section{Introduction} \label{s:intro} 
Due to advances in technology, it is now possible to have low-cost, stand-alone 
sensor and actuator devices 
that can communicate through the wireless medium. Such devices have applications in areas 
like epidemic detection, biological attack detection, intruder detection. In some
applications, these sensor 
nodes are deployed in hostile environments, and therefore
security of communication becomes critical.
To provide security, well-developed public key cryptographic methods have been considered, 
but these are compute-intensive and too  demanding for resource-constrained devices 
\cite{kruussymmetricreason}. So, symmetric key based encryption is the only way for 
secure communication between nodes. However, to do that, two nodes should agree upon a 
common key first. For this, various key distribution schemes have been proposed 
in the literature. Eschenauer and Gligor \cite{gligor} proposed a random key predistribution 
scheme, referred to as the basic scheme or EG scheme. Based on this scheme, various 
improvements have been proposed in the literature
\cite{jajodiarandumfunction,blom,varshneypairwiseblom,blundo-polynomialbased,blundopluseg-by-ning-polynom}.

Recently, there has been research on key distribution schemes which make use of  
predeployment knowledge of expected locations of the nodes, and these schemes are 
shown to perform better than the basic schemes. But in some cases predeployment knowledge 
is not available, so in \cite{groupbased} Liu et.al. proposed a scheme which uses 
group-based deployment model and showed that their scheme is better than the basic schemes.
Although in \cite{anjum-improvemen}, Anjum proposed a scheme which does not use  
predeployment knowledge of the nodes, and group-based deployment model, the scheme 
requires some nodes which can transmit at different power levels. 

Our scheme also uses the same concept of groups as used by 
\cite{varshneymain,groupbased,anjum-improvemen} but we have dropped the assumptions of 
predeployment knowledge of expected locations of nodes and group-based deployment. 
Moreover, our scheme does not require nodes which can transmit messages at different power 
levels.

\IEEEpubidadjcol

Our contributions in this paper are as follows.
\begin{itemize}
 \item We propose a scheme, in which nodes form groups \emph{after} 
deployment on the basis of their physical locations, and  generate keys which depend on the group 
they are part of. Each group is assigned a key pool and no two key pools have a common key. 
If a node is compromised, then it can affect communication in its group only; so the proposed scheme 
is more resilient to node capture than the basic schemes.
\item  We show using simulations that our scheme performs better than the basic scheme \cite{gligor} and the schemes which assume pre-deployment knowledge of node locations
\cite{varshneymain}. Our scheme assumes that there exists some constant time before which the 
adversary is unable to extract keys from the nodes, as assumed in \cite{leapjajodia,anjum-improvemen}. 
We show that this time can be considerably less for our scheme than that in \cite{leapjajodia}, with certain tradeoff.
\item We also look at the problem of connectivity of the key graph formed when our scheme is followed.
This problem is studied in the framework of  the ``AB random geometric graph'' \cite{iyer}.
Using results in \cite{iyer}, the number of tagged nodes is calculated such that the whole 
key graph is connected with high probability.
\end{itemize}

The rest of this paper is organized as follows. In Section~\ref{s:related_work}, we discuss related work. 
Section~\ref{s:proposal} describes our proposal. Section~\ref{s:abrandom} gives the expressions for the 
parameters ensuring connectivity of the key graph.
Section~\ref{s:sim_results} gives the comparison with the scheme proposed by Du.et.al \cite{varshneymain}. 
Section~\ref{s:otherapp} describes other applications where our scheme could be useful.
Section~\ref{s:conclusion} concludes the paper.

\section{Related Work} \label{s:related_work}
Various key distribution schemes have been proposed in the literature for wireless sensor networks, 
keeping in view the resource-constrained devices used in these networks. 
Eschenauer and Gligor \cite{gligor} proposed a scheme in which for every node, keys are
picked randomly (with replacement) from a key pool and assigned to it before deployment; 
this scheme is known as the basic or EG scheme. After key discovery,
two neighbor nodes that have a common key use that as the key for secure communication. 
Based on this basic scheme, several schemes with enhanced security features have been suggested 
in \cite{chanperrigrkpqcomp,varshneypairwiseblom,varshneymain,blundo-polynomialbased,anjum-improvemen}.

There is another class of schemes called ``threshold schemes.'' In these schemes, all nodes can 
communicate with one another, and no communication is compromised until some fixed number of nodes 
is compromised. Blundo et.al.\cite{blundo-polynomialbased} and Blom \cite{blom}  proposed such   
threshold schemes. Blundo's scheme uses symmetric bivariate polynomials to obtain pairwise keys, 
while Blom's scheme also uses a similar idea, in which symmetric matrices are used instead of 
symmetric polynomials. 

Du et. al. \cite{varshneypairwiseblom} improve upon Blom's scheme by combining it with the random 
key distribution scheme. Similarly,  Liu and Ning \cite{blundopluseg-by-ning-polynom} improve
upon Blundo's \cite{blundo-polynomialbased} scheme by combining it with random 
key distribution scheme.  Both these schemes perform better 
than the EG scheme \cite{gligor} in terms of connectivity and resilience 
against node capture. But threshold schemes do not scale with the number of nodes in the network. 
For a fixed resilience against node capture, if the number of nodes is increased, then 
they require large memory.

In any sensor network, generally nodes need to talk to their neighbor nodes only. So it is quite 
intuitive that nodes which are near should share the same key pool. This will lead to more 
efficient 
use of memory, and will give better connectivity and better resilience against node capture. 
Because of this reason, various location-based key distribution schemes have been proposed. 

Du et.al. \cite{varshneymain} and Liu and Ning \cite{ning-blom-2003} independently proposed 
schemes which assume predeployment knowledge of expected locations of the nodes. Nodes are assumed to 
be deployed in groups (group-based deployment model) and nodes in the same group have the same expected 
location, so that after deployment, they lie close to one another. Further in \cite{varshneymain}, 
nodes in the same group are allocated keys from the same key pool, while the groups which lie far 
from each other are allocated disjoint key pools. Therefore, compromise of any node jeopardises
transmissions of nearby nodes only. Due to this reason, performance is better than that of
the EG scheme.

All the location-based schemes which depend on the knowledge of expected locations of nodes 
perform well, but they are all prone to estimation errors in the expected positions of the nodes. 
So other schemes which do not assume predeployment knowledge of the expected locations of the 
nodes have been proposed. % but still their solution is location-dependent. 

In \cite{groupbased}, Liu et.al. proposed a scheme which does not use expected locations of the nodes 
but still uses group-based deployment. This scheme proposes a framework, and any basic scheme like 
random key distribution or polynomial-based scheme can be used with this framework. The authors 
showed that basic schemes used with their proposed framework perform better than when used alone.

Further, Anjum \cite{anjum-improvemen} removed the assumption of group-based deployment model and 
also removed the assumption of knowledge of expected locations of nodes. He showed that the 
scheme performs better than the basic scheme; but the scheme requires nodes which can transmit at different 
power levels. In this scheme there are some special nodes which generate different random 
numbers (nonces) and transmit them at different power levels. Nodes receiving the same nonce % map
% it to some different number. So all the nodes which receives the same nonce 
can communicate, 
provided they are neighbors. Our scheme is different from this scheme, since we do not require
the presence of nodes which can transmit at different levels. Instead of using different power 
levels, our scheme uses TTL scoping. In TTL scoping, after the deployment phase, some nodes 
transmit a broadcast packet containing the TTL (Time to Live) field, similar to that of IP 
packets in data networks. 

In addition, our scheme is also different in the way nodes choose their key rings. In 
\cite{anjum-improvemen}, on receiving the nonce, nodes map it to some different value. 
In contrast, in our scheme, some nodes transmit their id, and corresponding to every id there 
is an associated key pool. Nodes sample keys from the key pool corresponding to the received id. 
The main advantage of doing this is the improved resilience against node capture. % Since in the 
% case of 
In~\cite{anjum-improvemen}, all nodes receiving the same nonce use the same key for secure 
communication; so if any node is compromised, all the secure links formed by the nonce 
received by this node will be compromised. On the other hand, in our scheme,
communication with other nodes is compromised \emph{with some probability only},
because nodes receiving the same id sample keys from the key pool instead of using the same key.

\section{Proposed Scheme for Key Generation and Discovery} \label{s:proposal}
In this paper, we consider static sensor networks. Nodes are uniformly distributed across 
the deployment region. 
We use the following cryptographic primitives:
\begin{itemize}
\item Pseudo Random Number Generator (PRNG) --- This is a deterministic function, which takes an 
$n$ bit number as input and produces output of $m > n$ pseudorandom bits:
\begin{equation}
f :\{0, 1\}^n \rightarrow \{0, 1\}^m
\end{equation}

\item Hash function --- This is a deterministic function which takes an input of any length and 
returns a number of fixed bit-size. Given the output of the hash function, one cannot find the 
input and it is highly unlikely that for two different inputs, the output is same:
\begin{equation}
h:\{0,1\}^n \rightarrow \{0,1\}^c
\end{equation}
where $n$ is variable, and $c$ is fixed.

\end{itemize}

\subsection{Description of our scheme}
In our scheme, all nodes are divided into two sets: the ``tagged node'' set and 
the ``normal node'' set. Tagged nodes are similar to normal nodes in terms of memory, storage, 
and transmission range. They are deployed in the same way as normal nodes. 
The difference is that tagged nodes are 
programmed to broadcast a packet after the deployment phase is over.
Subsequently, tagged nodes behave like normal nodes.

Once deployment is over, a tagged node broadcasts a packet with TTL value $H$. Nodes within 
distance $Hr$ from the tagged node receive this packet, and all these nodes associate 
themselves in one group. Different groups are associated with \emph{disjoint} key pools, 
% (any two key pools do not have common keys), 
and nodes in a group ``sample'' keys from the same key pool. Since we are 
using disjoint key pools, so compromise of any node results in compromise of
communications in \emph{its} group only; in this way \emph{localization} of the effects of node
compromise is achieved. 

%As the groups are formed after the deployment of the nodes, so nodes will select the 
% key pools and sample keys \emph{after} the deployment of nodes. 
Our scheme relies on the assumption that the adversary 
will not be able to extract keying material from a captured
node before a small time interval has elapsed. This is a reasonable assumption because
breaking into a node and extracting keying material will take some time. The same assumption
has been made in \cite{leapjajodia} and \cite{anjum-improvemen}.

Our scheme consists of the following four phases:

\begin{itemize}
\item Predeployment Phase
\item Broadcast Phase
\item Key Generation Phase
\item Shared Key Discovery Phase
\end{itemize}

\subsubsection{Predeployment Phase}
In this phase, two keys are stored in the nodes. 
\begin{itemize}
\item Global key ($K_g$): This is common to all the nodes and is used for authentication and 
encryption of packets during the broadcast phase. 
\item Root Key ($K_r$): This key is a single key stored in all the nodes. It is used to derive 
the other keys during the key allocation phase; this procedure is explained subsequently. 
\end{itemize}
 
\subsubsection{Broadcast Phase}
After nodes are deployed, all tagged nodes broadcast a packet up to $H$ hops, containing two fields: 
Tagged node id field and Hop count field. Each node (both normal and tagged node) receiving this packet 
fetches the tagged node id from the packet, and compares it with previously stored tag ids. 
If there is no match with any of the previously stored tag ids, then its value is stored. 
Then, the hop count value is fetched from the packet and its value is decreased by $1$. 
If, after decreasing, the value is $0$, then the packet is discarded; otherwise, the packet is 
broadcast again with the new value of the hop count. This broadcast packet is encrypted and 
authenticated using the global key ($K_g$).  All nodes are able to decrypt and authenticate this 
packet since this key is stored in all the nodes. After the end of this phase, all nodes which 
are within distance $Hr$ from the tagged node receive the packet. In this way, all the nodes are 
divided into groups, on the basis of their physical locations.

Consider a tagged node $j$. It will broadcast a packet with tagged node id field set to $j$.
All nodes within the radius of $Hr$ of this tagged node will receive this packet and associate themselves with tagged node id $j$. All these nodes will consider 
themselves as a part of group $G_j$. So there will be a group 
corresponding to each tagged node. We note that, since a node can receive broadcasts from more than one 
tagged node, a node can be part of multiple groups.

There is a key pool corresponding to each group and each node samples $k$ keys from the key pool of each 
group to which it belongs. Since a node can be part of multiple groups, so different nodes can choose
different key ring sizes. To bound the number of keys chosen by any node, a limit is put on the number 
of key pools from 
which a node samples keys. Let $T_{key}$ be the maximum number of groups to which a node can belong.

We define two sets for any node $u$, $B_u$ and $T_u$. $B_u$ contains all the distinct tagged node 
ids received during the broadcast phase. $T_u$ is a subset of $B_u$. 
A selects $T_{key}$ tag ids out 
of the received tag ids, and the set $T_u$ contains these selected values.

Randomly selecting the tagged node ids from $B_u$ is not the best thing to do. Consider an example 
with $T_{key}=1$. If 
two neighbor nodes receive broadcasts from the same $4$ nodes, then on randomly selecting the tag id, 
the probability 
that both choose the same tagged node id is $1/4$. But if both the nodes plan to choose the least tag node id,
then with probability $1$ they will choose the same tag node id. And intuitively one can say that two neighbor 
nodes are more likely to receive broadcasts from the same set of tagged nodes. So we set the selection criterion 
as: Node $u$ selects the smallest $T_{key}$ tagged node ids from the set $B_u$.

\subsubsection{Key Generation Phase}

 Once the broadcast phase is over, nodes select $T_{key}$ smallest tag ids from the received tag ids. 
If we consider node $u$, then it is a part of groups in the set $G^u =\{G_j, \forall j \in T_u\}$. 
After the node has associated itself with the groups, it has to sample keys from the key pools 
corresponding to the selected groups. One way to do this is to store all the key pools in all 
the nodes before deployment, but this is not feasible because of memory constraints. So we propose a way 
in which nodes can \emph{compute} the keys such that it is \emph{equivalent} to first selecting the 
key pools and then sampling keys from them.

Let $P_j$ be the key pool associated with tag id $j$ or group id $j$.
The pool is generated by using \ref{eqn:key_pool}.
%All key pools are disjoint, which means no two key pools have a common key. If $k$ is the key ring size per group then node $u$ will sample $k$ keys from each key pool contained in set $P^u$ = $\{P_j, \forall\,\, j\in T_u\}$. Each group is also associated with unique Group Key $K_{K_{G_j}}$ and key pool $P_j$ each is generated using \ref{eqn:key_pool}

\begin{subequations}
\begin{gather}
P_j=\{h(K_{G_j}||i), 1\leq i \leq M\}\\
K_{G_j} = h^j(K_r)
\end{gather} \label{eqn:key_pool}
\end{subequations}

Here, 
$(||)$ represents the concatenation operator, $M$ is the key pool size per group, $K_r$ is the root key as defined earlier and 
$h^j(K_r)$ represents $j$ hash operations on $K_r$; for example $h^2(K_r) = h(h(K_r))$.

However, instead of deriving the key pool and then sampling the keys, one can first select 
$k$ numbers uniformly distributed in the range from $1$ to $M$, (called Key Indices), and then 
applying the function $h(K_{G_j}||i)$ to them. Each key can be identified by the tuple 
(Group Number, Key Index). Keys are stored along with this tuple to identify them during 
the subsequent key discovery phase.

To understand the procedure followed by nodes to derive their key rings, we consider a node $u$,
and examine what it does.

1) Node $u$ generates the set $K_u = \{K_{G_j}, \forall j\in T_u\}$, and arranges it in ascending order of its
 index values, $\{K_{G_{j_1}},K_{G_{j_2}}, ...\}$, with $j_2 > j_1$. Let us call these values ``Group Keys.'' If there are large number of tagged nodes then it will be expensive to compute the group key corresponding to tag nodes with large tag id's. So to minimize the computation, some fraction of group keys, uniformly distributed across the full range, can be computed offline and stored in the node 
before deployment. For example, if the number of tagged node is $1800$, then $36$ group keys 
($K_{G_{j_{50}}},K_{G_{j_{100}}}..$) could be pre-stored, so that the average number of 
hash computations done by any node will be $25$.

2) In this step, node $u$ generates $|T_u|$ sets each containing $k$ values in the range of $1$ to $M$. These sets are generated  using the PRNG, 
with $T_{key} u + 1, T_{key} u + 2 ... T_{key} u + |T_u| $ as the seed values. 
Let us call the elements of these sets ``Key Indices.''
 Since a node can generate a maximum of $T_{key}$ sets, so it will use $T_{key}$ seed values in the 
range $T_{key}u+1$ to $T_{key}(u+1)$. Further, nodes $u$ and $u+1$ will use different seed values since node 
$u$ will use seed values in the range $T_{key}u+1$ to $T_{key}(u+1)$, while node $u+1$ will use values in the range
$T_{key}(u+1)+1$ to $T_{key}(u+2)$. In this way, all sets of key indices are generated independently, and 
hence the key rings are also generated independently.

3) In this step, mapping of Key Indices to actual keys is done. Node $u$ has $|T_u|$ Group Keys and same number of 
Key Index sets. Node $u$ will pair each Key Index set with a single Group Key. Pairs are formed by first 
arranging the group keys in ascending order of their indices, and the Group Key with 
the smallest index is paired up 
with the Key Index set generated using $T_{key}u + 1$ as the seed value. 
From each pair, $k$ tuples are formed, where the first element of the tuple is the Group Key and the second 
element is the Key Index of the set. For example, suppose $T_u = \{2,5\}$. If the Key Index Set produced 
using $T_{key}u + 1$ is $\{1,9,10\}$ and that using $T_{key}u + 2$ is $\{11,91,56\}$,  
then following set of tuples is produced:\\
$K_{tup}$ = \{$(K_{G_2},1)$, $(K_{G_2},9)$, $(K_{G_2},10)$, $(K_{G_5},11)$, $(K_{G_5},91)$, $(K_{G_5},56)$\}
 
Note that each tuple can be identified by (Group Number($j$), Key Index). So even after the key $K_{G_j}$ is deleted from the memory of the node, tuples can be identified using the group number;
for the above example, tuples can be identified by  $\{(2,1),(2,9),(2,10),$ $(5,11),(5,91),(2,56)\}$.
This is important because during the key discovery phase, node will send it's node id ($u$) and the set 
$T_u$, and from this information, other nodes should be able to identify the common keys.

\begin{figure}[htbp]
\begin{center}
 \resizebox{8.5cm}{!}{\input{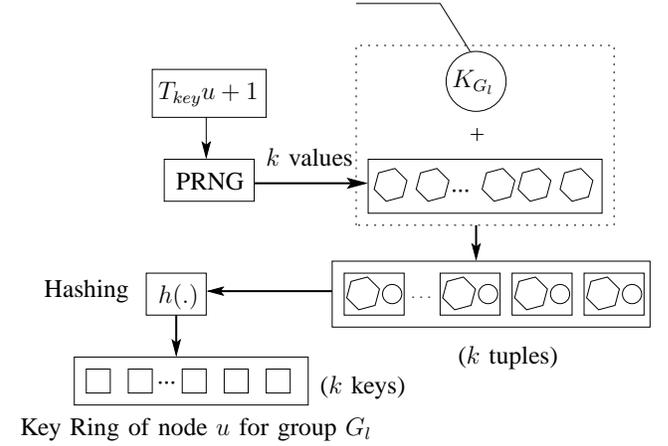}}
 \caption{Key Ring generation of node $u$ }
\label{figure:keygen1}
\end{center}
\end{figure}

Final keys are obtained by concatenating the elements of the tuple and then hashing the resultant value. This procedure is also illustrated in Fig. \ref{figure:keygen1}.

Our scheme requires nodes that are close to each other to be in the same group, nodes in the same group to 
sample keys from the same key pool and key pools selected by distinct groups to be disjoint. 
The procedure described above satisfies all our requirements. 

As soon as the key allocation phase is over, $K_r$ and the set $K_u$ should be 
deleted from memory, because given this information, an
attacker may be able to generate all the keys in the network, and that will lead 
to compromise of all communication. 

\subsubsection{Shared key discovery phase}
 In this phase, node $u$ broadcasts its node id and the set $T_u$. If node $v$ is a neighbor of node $u$, 
then node $v$ will receive the broadcast by node $u$. On receiving this broadcast packet, node $v$ will 
fetch the set $T_u$ and then compare its elements with $T_v$. If there is no matching element between the
two sets, then there is no common key. If there are common elements between the two sets, then call the set of 
matching elements as $T_{uv}$, $T_{uv}=T_u \cap T_v$. Let us call an element of $T_{uv}$ as $t_{uv}$. 
Now node $v$ will find the index of the elements present in the set $T_{uv}$, in sets $T_u$ and $T_v$.
What is done with the index is explained using the example below.

For example, let $T_u=\{1,5,9,13\}$ and $T_v=\{1,4,13,15\}$, so the set $T_{uv}=\{1,13\}$, and 
the index of these elements in set $T_u$ is $1$ and $4$, while in set $T_v$ it is $1$ and $3$.

Now node $v$ will generate $k$ numbers for seed values $T_{key}u + 1$, $T_{key}u + 4$ and $T_{key}v + 1$, 
$T_{key}v + 3$ using the PRNG. Then it will compare the values produced from seed values $T_{key}u + 1$ and 
$T_{key}v + 1$, and also compare the values produced from the seed values $T_{key}u + 4$ and $T_{key}v + 3$.

If any value matches, then the node $u$ and $v$ share a common key, and as mentioned earlier, the key is 
identified by the tuple (Group Number, Key Index). 
If multiple keys are found to be shared, then XOR of all the keys will be used as the common key. 
On following a similar procedure, node $u$ can also find common keys with node $v$.

\section{Design Parameters ensuring Connectivity} \label{s:abrandom}

After keys have been generated and discovered, the natural question which arises
is: Can any two nodes exchange information securely? This question is addressed
by considering the notion of the key graph and connectivity of the key graph.

A node is represented by a vertex in the graph.
There exists an edge between two vertices if the corresponding nodes share at least one common key, and 
lie in the coverage radius of each other. A graph formed in this way is called a ``key graph.'' 
There are two properties of connectivity: local connectivity and global connectivity. 
Local connectivity of any node is the probability of sharing at least one key 
with the neighbor nodes, while global connectivity is the percentage of the nodes in the key  graph
that is reachable from any node. 

In this section, we are concerned with the problem of
how to find the number of tagged and normal
nodes such that the resulting key graph is connected. The
problem of connectivity of the key graph can be broken down into followingthe  three sub problems: 
\begin{itemize}
\item IntEr Group connectivity (IEG) --- 
Since key pools are disjoint, so nodes belonging to two different groups will have zero probability 
of sharing a key. So, two groups can only be connected if there exist $l$ nodes ($ l \geq 1$), 
which belong to both the groups. As these common nodes sample keys from the key pools of both the 
groups, so these nodes are reachable from the nodes of both the groups; 
thus, these nodes act as ``gateways.''

Now consider a graph in which a \emph{group} is represented by a vertex, and there exists an edge 
between two vertices, if there exists at least one common node between two groups. 
In \emph{this} graph, if all the vertices are reachable from any vertex, then the IEG property holds. 
We will use the results in \cite{iyer} %Iyer and Yogeshwaran 
on AB random geometric graphs to find the minimum number of groups or tagged nodes 
required such that all the groups are connected.

In \cite{iyer}, two kinds of nodes are considered: A type and B type. 
Two A type nodes can communicate via a B type node only. 
Let the graph formed in this way \emph{among A type nodes}, be denoted as $G(n,cn,r_n)$,
where $n$ is the number of A type nodes and $cn$ is the number of B type nodes. 
We apply this framework by taking type A nodes as tagged nodes, and type B nodes as
normal nodes.

Define $M_n$ as the largest nearest neighbor radius of the AB random geometric graph, i.e., 
the radius below which there exists at least one node with degree equal to zero. 
Then, \cite{iyer} shows that 
$\lim_{n \to \infty }P(M_n\leq r_n) = e^{-\beta}$ for transmission radius ($r_n$) 
equal to $\sqrt{\frac{log(n/\beta)}{cn\pi}}$. 

Also, the thereshold transmission radius of the nodes for which the graph $G(n,cn,r_n)$
is connected with high probability as $n \to \infty$ is given by Eqn. \ref{eqn:iyer_radius}:
\begin{equation}
 r_n = \left(\frac{2 + \sqrt{c}}{2}\right)\left(\frac{log(n/\beta)}{cn\pi}\right)^{.5}
 \label{eqn:iyer_radius}
\end{equation}
We will fix $\beta$, which will translate to a target small probability
of ``graph isolated'' groups. 
Then, we find the value of number of tagged nodes ($n$) required, 
for the given value of $r_n=Hr$ and total number of nodes to be deployed ($N = n(1+c)$), 
by substituting $r_n = Hr$ and $c=\frac{N-n}{n}$ in Eqn. \ref{eqn:iyer_radius}. If $n^*$ is the solution obtained  and if $T_{i} = \lceil n^* \rceil$, then the number of tagged nodes greater 
than $T_{i}$ will satisfy IEG property.

\item Tagged node-covered --- Nodes which do not receive broadcasts will not be part of any group;
so these nodes are isolated from rest of the network. The
number of tagged nodes should be such that all the normal nodes are covered by broadcast from at least 
one tagged node. We will use the result from \cite{zhanaghoulifetime} to calculate the expected 
number of normal nodes not covered by any broadcast. It is given by Eqn. \ref{eqn:kcovered0}.
\begin{equation}
E[N_{t}] =(N-n) e^{-\frac{n}{A} \pi (Hr)^2}
\label{eqn:kcovered0}
\end{equation}
where, $A$ is the deployment area.
To satisfy this property, $E[N_{t}]$ should be less than $1$. If $n^*$ is the solution of the equation 
$E[N_{t}] = 1$, and if  $T_{c} = \lceil n^*\rceil$, then the number of tagged nodes greater 
than $T_{c}$ will satisfy node-covered property.

So, to satisfy both the node-covered property and and inter group connectivity, the number of tagged nodes
is given by Eqn. \ref{eqn:calcT}.
\begin{equation}
T>max(T_{c},T_{i}) \label{eqn:calcT}
\end{equation}

\item IntrA Group connectivity (IAG) --- All the nodes within a group should be reachable from any 
node in the key graph. We will ensure this by using expressions to 
calculate the keyring size
in the EG scheme \cite{gligor}.  Since the number of nodes in a group has decreased, so less number of 
nodes will share the key pool. So, key pool size could be reduced, which, in turn, will reduce the 
requisite keyring size.
\end{itemize}

If all the above mentioned requirements are met, then our objective of connectivity of the 
key graph is achieved, because all the nodes are reachable within a group, all the groups are 
reachable from any group and all the nodes are part of at least one group.

\section{Evaluation \& Comparison}\label{s:sim_results}

We compare our proposed scheme with the scheme~\cite{varshneymain} which makes the stronger assumption of availability of expected knowledge of positions of nodes, which our scheme does not. However our scheme  makes another assumption: that there exists an interval (vulnerable time), after the deployment, during which attacker should not be able to extract the keys. 
It may be noted that in our scheme, keys are generated by nodes 
\emph{after} they are deployed, while in \cite{varshneymain}, keys are loaded into nodes
before deployment. Moreover,
our scheme also has the features of random key distribution. So it is appropriate to compare 
the connectivity and resilience of our scheme with that of schemes which use random key 
pre distribution (RKD) schemes 
(for example, \cite{gligor}) and a scheme like in \cite{varshneymain}, which uses RKD with 
the assumption of knowledge of expected locations of the nodes. We will also compare our scheme 
with LEAP$+$ in \cite{leapjajodia}, which is also a post-deployment key generation scheme
and considers the notion of vulnerable time.
We will argue that our scheme leads to a smaller vulnerable time than that required by 
in LEAP$+$, and discuss a related trade-off.

For the evaluation of our scheme, we use the following metrics.
\begin{itemize}
\item Connectivity of the nodes in the key graph: local and global connectivity.
\item Resilience against node capture --- This is the ratio of the number of links compromised to 
the total number of links formed. Links which are formed among the compromised nodes and the links 
which are between the compromised nodes and the noncompromised nodes are not taken into account 
while calculating the number of compromised links and the total links.
\item Memory requirement --- As the sensor nodes are resource-constrained, so there is always the 
requirement of attaining high connectivity using minimum memory.
\item Vulnerable Time --- It is the time interval after the deployment during which attacker should not be able to extract key material from the nodes, and before which all nodes should delete the key material from their memory. It is desirable to keep it as less as possible.
\end{itemize}

Our simulations are done for the following values of parameters. 
The number of tagged nodes is calculated 
using the analysis in the previous section.
\begin{itemize}
\item Deployment region: $1000 m \times 1000 m$
\item Total number of nodes $(N)$ is taken as $10000$
\item Transmission radius of the node is taken as $40 m$
\item Hop Count ($H$) is taken as $1$
\item $T_{key}=2,4$, Keys per key pool $M = 1000$
\item Tagged nodes: $T_i = 1863, T_c = 1794$, $T = 1863$ 
\end{itemize}
Also, simulations for the scheme \cite{varshneymain} are done using the same set of values. 
\subsection{Simulation Results - Connectivity}

\begin{figure*}[!t]
\subfloat[Comparison with other schemes]{\includegraphics[width
= 3.5 in]{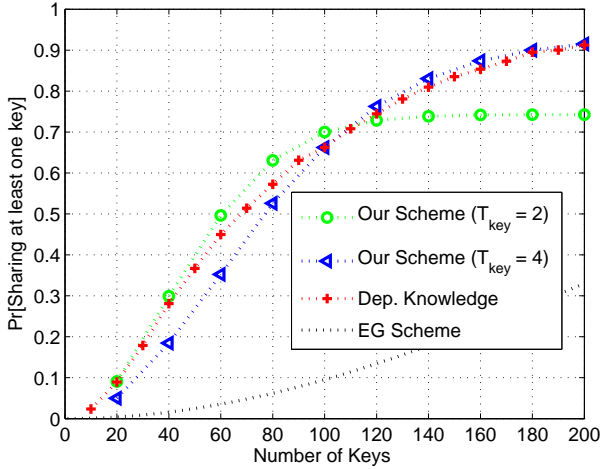}
\label{fig_all_1}} 
\hfil
\subfloat[Analysis Vs Simulation (of our proposed scheme)]{\includegraphics[width
= 3.5 in]{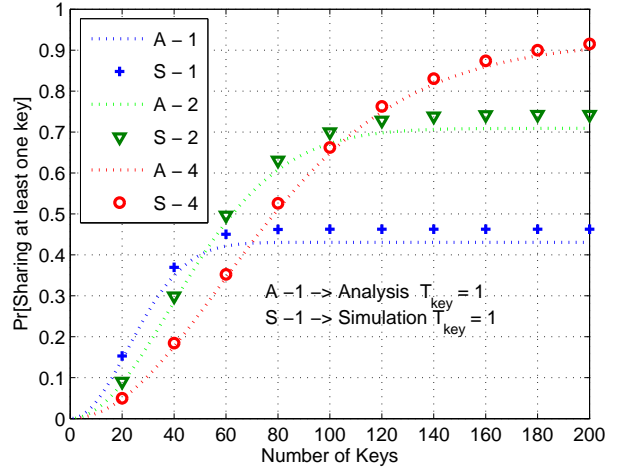}
\label{fig_sim_anal_h2}} 
\caption{Connectivity Analysis}
\end{figure*}

\subsubsection{Local Connectivity}
Figure~\ref{fig_all_1} shows the plot of probability of sharing at least one key between two neighbor nodes versus the key ring size of our scheme, scheme \cite{varshneymain} and EG scheme. We have simulated our scheme for $T_{key} = 2,4$.
From the graph it is clear that our scheme performs better than or as well as the scheme \cite{varshneymain}. For lower values of key ring size, $T_{key} = 2$ should be chosen which gives better connectivity than that of the scheme \cite{varshneymain}, while for larger value of key ring sizes $T_{key} = 4$ should be chosen, which gives same connectivity as that of the scheme \cite{varshneymain}. 
We have derived analytical expressions for the probability of sharing at least one key between two 
neighbor nodes, and the results match well with the simulated values as shown in 
Fig.~\ref{fig_sim_anal_h2}. 
The analytical expressions are not reported here due to lack of space.

%\begin{figure*}[t]
%\centering
%\subfloat[Comparison with other schemes]{ \includegraphics[width =2 in]{hop23tkey2all}\label{fig_all_1}}
%\hfill
%\subfloat[Analytical Vs Simulation]{\includegraphics[width=2 in]{hop2theoryvssimulation}
%\label{fig_sim_anal_h2}}
%\caption{Local Connectivity}
%\end{figure*}

%\begin{figure}[!t]
%\centering
%\includegraphics[width=2 in]{hop23tkey2all}
%\caption{Local Connectivity: Comparison with other schemes:}
%\label{fig_all}
%\end{figure}

\begin{figure*}[!t]
\subfloat[For Local conn. ($p$) $= 0.33$]{\includegraphics[width
= 3.5 in]{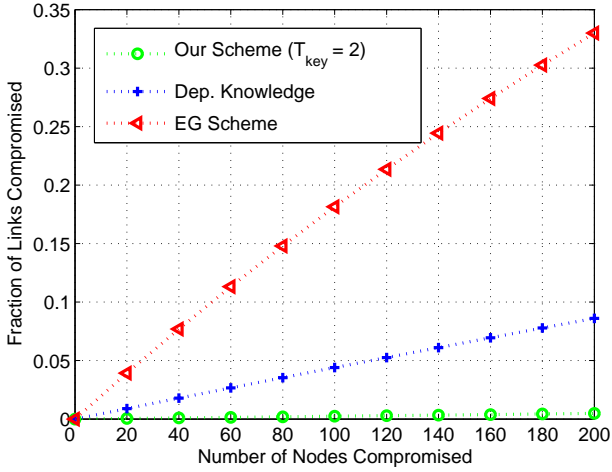}\label{fig_security33}} 
\hfil
\subfloat[For Local conn. ($p$) $= 0.50$]{\includegraphics[width=3.5 in]{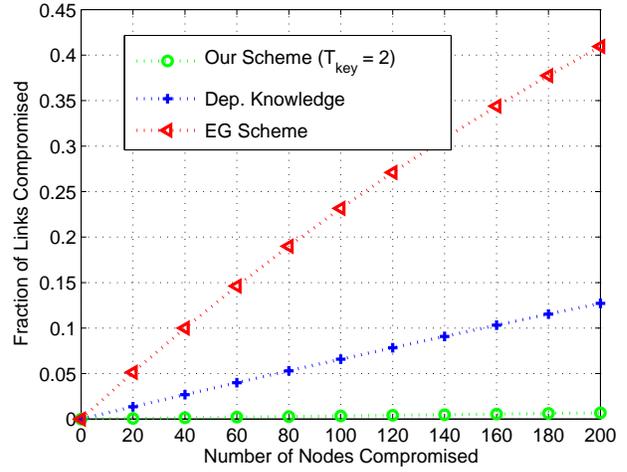}
\label{fig_security50}}
\caption{Security Analysis}
\end{figure*}

\subsubsection{Global connectivity}
For $T_{key} = 2,4$, Table~\ref{tab:locvsglob} gives the comparison of local connectivity
with global connectivity of the nodes. 
As we can see from the last row of the table,
there are small percentages ($.08$ and $.12$) of nodes which are isolated even at 
large value of key ring size. This is because they do not receive broadcasts from any of the 
tagged nodes, and hence remain isolated from the network.

However, an algorithm could be developed to tackle this problem of a small number of nodes not receiving broadcast from any tagged node. The simplest algorithm is as follows: All nodes which do not receive a broadcast from 
any tagged node 
send a packet containing their node id to the sink node, and broadcast it up to one hop. Nodes which receive 
\emph{this} broadcast packet fetch the node id from the packet, and send a message to the sink node 
(authenticated and encrypted with their own key) containing the node id they received from the 
broadcast packet and their group id. The sink node samples keys from the group key pool 
of the neighbor nodes and communicates it to the uncovered node.

%\begin{table}[!t]
%\renewcommand{\arraystretch}{1.3}
%\caption{Local Vs Global Connectivity($H=2,T_{key}=2$)}
%\label{tab:locvsglob}
%\centering
%\begin{tabular}{|c|c|c|} \hline
% Number of Keys ($2k$) & Local  & Global \\ \hline%
%			& &		   \\ \hline
% 		20     & 0.0653 &  99.99  \\
%      		60     & 0.1400 &  99.4300 \\
%      		100    & 0.7463 &  99.9950  \\ \hline
%\end{tabular}
%\end{table}
 
\begin{table}
\renewcommand{\arraystretch}{1.4}
\caption{Local Vs Global Connectivity}
\label{tab:locvsglob}
\centering
\begin{tabular}{|c|c|c|c|c|}
\hline
No. of Keys&\multicolumn{2}{c|}{Local}&\multicolumn{2}{c|}{Global} \\
\hline
($T_{key}k$)&\multicolumn{2}{c|}{$T_{key}$}&\multicolumn{2}{c|}{$T_{key}$} \\\hline
    & $2$ & $4$   & $2$  & $4$ \\\hline
$40$  & $0.29$ & $0.18$ & $99.63$ &  $99.71$ \\\hline
$60$  & $0.49$ & $0.35$ & $99.84$ & $99.92$   \\ \hline
$100$ & $0.69$ & $0.66$ & $99.88$ &  $99.92$    \\ \hline
\end{tabular}
\end{table}

\subsection{Security Analysis}
The basic threat model that we have considered is as follows:
If a node is captured, then all the keys contained in it are revealed to the adversary.
To evaluate the resilience of our scheme, we consider that $x$ nodes are compromised and they are 
distributed uniformly across the deployment region. Key ring size of any node is $L = T_{key}k$, where 
$k$ is the key ring size per group. Total number of keys in the key pool is given by  $S = TM$, where 
$M$ is key pool size per group, and $T$ is number of tagged nodes.
  If one node is captured, then the probability that a link between two non-captured nodes
is compromised is at most
$\frac{L}{S}$. (This is a  
worst case value since links between non-captured nodes could be secured by multiple keys).
$\frac{L}{S}$. For $x > 1$ compromised nodes, the
probability that a link is not compromised is at least
$\left(1 - \frac{L}{S}\right)^x$. Then, the probability that a 
link is compromised is at most $1 - \left(1 - \frac{L}{S}\right)^x$. While comparing 
this metric with other schemes, local connectivity ($p$) needs to be kept the same. 
Figure \ref{fig_security33}, \ref{fig_security50}
show a comparison of the resilience metric of our scheme with that of other schemes, 
at local connectivity $p=0.33$ and $p=0.5$.
 It shows that our scheme performs better than other schemes. We see that the fraction of links 
compromised due to node capture is very small compared to that in \cite{varshneymain}. 
This is attributed to the fact that our scheme has small group size, which is equal to the 
total number of nodes falling in the transmission radius of the node. So when a node is 
compromised, it affects communication links of the groups it is part of, and these
are very small in size; so the fraction of total links compromised is also very small. 
In \cite{varshneymain}, the group size is bigger, so the number of affected nodes is larger, 
and so is the fraction of compromised links. However, it could be argued that in this scheme 
also the number of groups can be increased, simultaneously decreasing the number of nodes a group; 
but this requires expected locations to be known with more precision, with a corresponding 
increase in the complexity of deployment. 
\subsection{Vulnerable Time}
Since in our scheme keys are computed after the nodes are deployed, so it is important to consider another threat, which is the vulnerable time during which the compromise of a single node can lead to compromise of whole network. We will compare our scheme with LEAP$+$, which also needs to 
address the issue of vulnerable time.
Since we have shown that our scheme for ($H=1$) performs better than the schemes using deployment knowledge, so we will compare the vulnerable time for $H=1$ only. Vulnerable time required by our scheme for ($H=1$) is just the time taken by all the tagged nodes to broadcast up to one hop or time taken by all the nodes to hear broadcast from all the neighbor tagged nodes;
the latter constitute a fraction ($.18$) of the total number of nodes.
In LEAP$+$ ~\cite{leapjajodia}, it is the time taken by all the nodes to hear broadcast from all the neighbor nodes. Since the number retransmission attempts required by the node to transmit in any MAC protocol depends on the number of active neighbor nodes, so our scheme, which requires that nodes should hear from  only neighbor tagged nodes which are only the fraction of total number nodes, requires less time, than what LEAP+ does. However, there is a trade off between performance achieved 
and the requirement of the vulnerable time, since LEAP+ acheives perfect resilience and 
connectivity while our scheme does not. Nevertheless, still our scheme is able to achieve better performance than the schemes which assumes predeployment knowledge of the nodes.
\subsection{Addition of New Nodes}
 To add new nodes after the initial deployment, the base station or sink informs tagged nodes to broadcast their tag id's for the new nodes. Root key is stored in the new nodes before their deployment, using which they generate the keys. Key generation and shared key discovery procedure is same as described earlier. After the generation of the key ring they delete the root key.

%\begin{figure*}[!t]
%\centerline{\subfloat{\includegraphics[width =2 in]{security33}}
%\caption{Security Analysis : Comparison with other schemes (a) $p = 0.33$ (b) $p = 0.5$}
%\label{fig_security33}
%\hfil
%\subfloat{\includegraphics[width=2 in]{security50}}}
%\caption{Security Analysis : Comparison with other schemes (a) $p = 0.33$ (b) $p = 0.5$}
%\label{fig_security50}
%\end{figure*}

%\begin{figure*}[!t]
%\centering
%\subfloat[For Local connectivity($p$) $= 0.33$]{\label{fig_security33} \includegraphics[width =2 in]{security33}}
%\hfill
%\subfloat[For Local connectivity($p$) $= 0.50$]{\label{fig_security50}\includegraphics[width=2 in]{security50}}
%\caption{Security Analysis}
%\end{figure*}

%\begin{figure*}[!t]
%\centerline{\subfloat[For Local connectivity($p$) $= 0.33$]\includegraphics[width=2.5in]{./eps/security33}
%\label{fig_security33}}
%\hfil
%\subfloat[For Local connectivity($p$) $= 0.50$]{\includegraphics[width=2.5in]{security50}
%\label{fig_security50}}}
%\caption{Security Analysis}
%\label{fig_sim}
%\end{figure*}

\section{Other Applications}\label{s:otherapp}
Apart from the main objective of key distribution, our scheme could be used for other applications as well.
Tagged nodes could be considered as the virtual base stations (VBS) distributed all across the deployment 
region, and other nodes around them associated with them. 
If the main base station wants to convey a broadcast message only to some nodes lying in a certain region, 
then this could be accomplished by a single unicast message to the tagged node
(assuming tagged nodes can identify their location) lying near that region and then that tagged node 
can broadcast that message with a flag indicating a ``regional broadcast,'' so that receiving nodes
check the hop count value before rebroadcasting the message. All the tagged nodes can maintain a 
group key which is known only to its group members; this will help secure delivery of the message.  

This scheme can also be used for group key management. In a distributed environment where all nodes 
cannot communicate directly with base station, tagged nodes can act as the virtual base stations. 
Such a scheme involving decentralization is proposed in \cite{kct_cluster}; however, in this scheme, cluster 
heads (tagged nodes) have larger transmission range than the cluster members (normal nodes), 
so that they can communicate directly with sink nodes, and are less energy-constrained than the normal nodes. 
Another application could be the collection of data, where the nodes in the group sends their data to the 
VBS of their respective group, and then VBS creates a single packet and send it to BS.

\section{Conclusion}\label{s:conclusion}
We have proposed a key distribution scheme, which does not assume node predeployment knowledge and also 
does not require nodes to be deployed in groups, still our scheme achieves better performance in terms 
of connectivity and security than the scheme~\cite{varshneymain} which take these assumptions.  However our scheme assumes existence of vulnerable time, which is less than that of the LEAP+~\cite{leapjajodia}, with tradeoff of connectivity and resilience.
\newpage
Our future work is to propose a way to allocate polynomial based keys~\cite{blundo-polynomialbased}, 
to make our scheme more robust against node capture attack. Another future work is to evaluate and compare 
the performance of our scheme for non-uniform deployments.
%

%\bibliographystyle{plain}
%\bibliography{sigproc}  %
%\bibliographystyle{IEEEtran}
%\bibliography{IEEEabrv,mybibfile}
%\bibliography{sigproc}

\section{Acknowledgments}\label{s:acknowledgements}
    This work was supported by DRDO under Project
     571: A Research and Implementation Project on Sensor 
     Networks.

\bibliographystyle{abbrv}

\end{document}